\documentclass[aps,prd,twocolumn,floatfix,preprintnumbers,superscriptaddress]{revtex4-1}
\usepackage{xcolor}

\usepackage[margin=0.7in]{geometry}
\usepackage[utf8]{inputenc}
\usepackage[english]{babel}
\usepackage[normalem]{ulem}
\usepackage{amsmath,amssymb,amsthm,amsfonts,mathtools}
\usepackage{graphicx}
\usepackage{subcaption}
\usepackage{enumitem}
\usepackage{indentfirst}
\usepackage{textcomp}
\usepackage{multirow}
\usepackage{slashed}
\usepackage{braket}
\usepackage{physics}
\usepackage{appendix}
\usepackage{url}
\usepackage{natbib}
\usepackage{mathrsfs}
\usepackage{cancel}
\usepackage[normalem]{ulem}
\usepackage{array}
\usepackage{booktabs}
\usepackage{verbatim}
\usepackage{ragged2e}
\usepackage{soul}
\usepackage{xcolor}
\usepackage{hyperref}
\usepackage{cleveref}
\usepackage{orcidlink}

\hypersetup{
    colorlinks=true,
    linkcolor=blue,
    citecolor=blue,
    filecolor=blue,
    urlcolor=blue
}

\usepackage{tikz}
\usetikzlibrary{positioning,decorations.pathmorphing,decorations.markings,arrows}
\usepackage[compat=1.1.0]{tikz-feynman}

\newcommand*\circled[1]{\tikz[baseline=(char.base)]{
            \node[shape=circle,draw,inner sep=1pt] (char) {#1};}}

\begin{document}

\begin{flushright}
MI-HET-855
\end{flushright}

\title{Photon Excess from Dark Matter and Neutrino Scattering at MiniBooNE and MicroBooNE}

\author{Bhaskar Dutta~\orcidlink{0000-0002-0192-8885}}
\email{dutta@tamu.edu}
\affiliation{
Mitchell Institute for Fundamental Physics and Astronomy,
Department of Physics and Astronomy, Texas A\&M University, College Station, TX 77845, USA
}
\author{Aparajitha Karthikeyan~\orcidlink{0000-0002-6934-1239}}
\email{aparajitha\_96@tamu.edu}
\affiliation{
Mitchell Institute for Fundamental Physics and Astronomy,
Department of Physics and Astronomy, Texas A\&M University, College Station, TX 77845, USA
}
\author{Doojin Kim~\orcidlink{0000-0002-4186-4265}}
\email{doojin.kim@usd.edu}
\affiliation{Department of Physics, University of South Dakota, Vermillion, South Dakota 57069}

\author{Adrian Thompson~\orcidlink{0000-0002-9235-0846}} 
\email{a.thompson@northwestern.edu}
\affiliation{Northwestern University, Evanston, IL 60208, USA}

\author{Richard G. Van de Water~\orcidlink{0000-0002-1573-327X}}
\email{vdwater@lanl.gov}
\affiliation{Los Alamos National Laboratory, Los Alamos, NM 87545, USA}


\begin{abstract}
We propose new solutions to accommodate both the MiniBooNE electron-like and MicroBooNE photon low-energy excesses, based on interactions involving light dark matter and/or neutrinos. The novelty of our proposal lies in the utilization of a photon arising from 2-to-3 scattering processes between a nucleus/nucleon and a neutrino and/or dark matter via exchanges of light mediators. We find that viable regions exist in the coupling and mass parameter space of the mediators and light dark matter that can simultaneously explain the observed excesses and remain consistent with current experimental constraints. We further highlight that these scenarios can be tested with upcoming data from various ongoing experiments. 
 \end{abstract}

\maketitle

\noindent {\bf Introduction.}
The recent MicroBooNE observation of a (mild) single-photon excess
at $2.2\sigma$~\cite{MicroBooNE:2025ntu} lends support to the earlier MiniBooNE observation of excess electron-like events at $4.8 \sigma$~\cite{MiniBooNE:2008yuf, MiniBooNE:2018esg, MiniBooNE:2020pnu}. Addressing this excess necessitates new physics beyond the Standard Model (SM), which is already motivated by unresolved questions such as the origin of dark matter, neutrino masses, etc. The excess thus presents a valuable
opportunity to investigate the potential roles for dark matter or neutrino interactions, which can be further examined in ongoing short-baseline neutrino experiments such as SBND~\cite{MicroBooNE:2015bmn} and ICARUS~\cite{ICARUS:2004wqc}. 

Existing solutions to this anomaly mostly involve additional species in the neutrino sector, 
utilizing their oscillation, decays, and/or upscattering ~\cite{Sorel:2003hf,Karagiorgi:2009nb,Collin:2016aqd,Giunti:2011gz,Giunti:2011cp,Gariazzo:2017fdh,Boser:2019rta,Kopp:2011qd,Kopp:2013vaa,Dentler:2018sju,Abazajian:2012ys,Conrad:2012qt,Diaz:2019fwt,Asaadi:2017bhx,Karagiorgi:2012kw,Pas:2005rb,Doring:2018cob,Kostelecky:2003cr,Katori:2006mz,Diaz:2010ft,Diaz:2011ia,Gninenko:2009ks,Gninenko:2009yf,Bai:2015ztj,Moss:2017pur,Bertuzzo:2018itn,Ballett:2018ynz,Fischer:2019fbw,Moulai:2019gpi,Dentler:2019dhz,deGouvea:2019qre,Datta:2020auq,Dutta:2020scq,Abdallah:2020biq,Abdullahi:2020nyr,Liao:2016reh,Carena:2017qhd,Abdallah:2020vgg, Abdallah:2024uby, Hammad:2021mpl}. Recently, long-lived mediators and dark matter upscattering have been considered to solve the MiniBooNE puzzle~\cite{Dutta:2021cip,PhysRevD.109.095017}. 
MicroBooNE's strong capability to distinguish photons from electrons~\cite{MicroBooNE:2025khi} 
has already constrained 
the parameter space of some of these solutions, in particular, through a potential signature in the inclusive single-photon channel~\cite{MicroBooNE:2025ntu} but without any obvious excess in the coherent-photon production channel similar to SM~\cite{MicroBooNE:2025rsd,Hagaman:2025mrx}. Already, these hints are beginning to inform a better understanding of the anomaly as it takes shape. 

In this work, we propose a novel explanatory path that leverages single-component dark matter and/or the three known SM neutrino species.
Our solutions involve a single-photon final state arising from neutral-current (NC) processes of the SM neutrinos $\nu$ {\it or} a 
new light fermionic dark matter\footnote{We do not investigate the origin of dark matter, as this is unrelated to the mechanism we propose and lies beyond the scope of the present study.} 
through $\nu \mathcal{N} \to \nu \mathcal{N} \gamma$ or $\chi \mathcal{N} \to \chi \mathcal{N} \gamma$ respectively, with $\mathcal{N}$ denoting a nucleus/nucleon.
Among other explanations for the anomaly, single-photon production through NC interactions in neutrino scattering was initially proposed 
in Refs.~\cite{Harvey:2007rd,Harvey:2007ca}. 
This mechanism manifests through $\nu \mathcal{N} \to \nu \mathcal{N} \gamma$, arising from a Wess-Zumino-Witten (WZW) anomaly accompanying the SM $Z$, photon, and the $\omega$ vector meson, which appeared to enjoy a large coupling. This was revisited and understood better in the context of the spontaneously broken local hidden symmetry formalism~\cite{Harada:2011xx}. However, it was later recognized that this channel's contribution to the NC single-photon production rate is too small---smaller than contributions from $\Delta$ resonances~\cite{Hill:2009ek,Rosner:2015fwa} (see also~\cite{Alvarez-Ruso:2014bla,Acero:2022wqg}). Nevertheless, the topology of this $2 \to 3$ scattering process and its resulting signature offer valuable insights that can help guide a new interpretation of the excess. In our scenario, $\nu \mathcal{N} \to \nu \mathcal{N} \gamma$ or $\chi \mathcal{N} \to \chi \mathcal{N} \gamma$ are not mediated through WZW, but instead by new dimension-5 operators that connect the SM photon to a new massive scalar and massive vector boson. 
The Yukawa couplings of quarks to the scalar, along with the coupling of neutrinos or dark matter to the vector mediator, enable the $2\to 3$ scattering.

Neutrinos and dark matter can arise from charged meson two- and three-body decays~\cite{Dutta:2021cip}, correlating their fluxes with the MiniBooNE target-mode excess via magnetic focusing in the BNB horn system.
This mechanism simultaneously accounts for the absence of an excess in MiniBooNE's dump-mode observations~\cite{MiniBooNEDM:2018cxm}, where the fluxes are isotropically suppressed due to the lack of  
magnetic focusing horns~\cite{Dutta:2021cip}.

\begin{figure}
    \centering
    \begin{tikzpicture}
        \begin{feynman}
            \tikzset{every node/.style={font=\large}}
            \vertex (i) at (-2, 1.1) {$\nu/\chi$};
            \vertex (f) at (2, 1.1) {$\nu/\chi$};
            \vertex (Ni) at (-2,-1.1) {$\mathcal{N}$};
            \vertex (Nf) at (2,-1.1) {$\mathcal{N}$};
            \vertex (if) at (0, 1.1);
            \vertex (Nif) at (0, -1.1);
            \vertex (c) at (0,0);
            \vertex (g) at (1.8,0) {$\gamma$};
    
            \diagram*{
                (i) -- [fermion, very thick] (if) -- [fermion, very thick] (f),
                (Ni) -- [double, thick] (Nif) -- [double, thick] (Nf),
                (if) -- [boson, very thick, edge label=$Z^\prime$] (c),
                (c) -- [scalar, very thick, edge label'=$\phi$] (Nif),
                (c) -- [photon, very thick] (g),
            };
        \end{feynman}
    \end{tikzpicture}
    \captionsetup{justification=Justified}
    \caption{Feynman diagram depicting the scattering of dark matter/neutrinos off a nucleus/nucleon $\mathcal{N}$ via the $Z^\prime$ and $\phi$ exchange to produce a single photon.}
    \label{fig:feynman_scattering}
\end{figure}

\medskip

\noindent {\bf Model.} 
We consider a simplified model which includes a massive scalar coupling to quarks, and a massive vector boson couples to either neutrinos, dark matter, or both simultaneously. Under this scenario, the scattering process delineated in Fig.~\ref{fig:feynman_scattering}
is expected to result in a single photon. 
A similar process occurs in the SM with neutrinos, via the $Z$-$h$-$\gamma$ interaction with the SM $Z$ and Higgs $h$. However, the associated cross section and kinematics from these heavy mediators are unsuitable to explain the MiniBooNE excess. 
This motivates us to introduce
a new gauge boson, $Z^\prime$, and a scalar, $\phi$, {\it lighter} than the $Z$ and $h$. The new scalar $\phi$ can emerge from the extensions of the SM Higgs sector~\cite{Dutta:2020scq, Dutta:2022fdt}, or as a new Higgs field in association with spontaneous breaking of the new $U(1)$ symmetry~\cite{Dutta:2019fxn}.
The effective Lagrangian for $\phi$ and $Z^\prime$ with masses $m_{\phi}$ and $m_{Z^\prime}$, respectively, relevant for the $2\to 3$ neutrino scattering, is
\begin{equation}
    \mathcal{L}_\nu\supset g_\nu \bar{\nu}_\ell\gamma^\alpha\nu_\ell Z_\alpha'-\frac{g_{\phi Z^\prime \gamma}}{2}\phi F_{\mu\nu}{Z^\prime}^{\mu\nu}+ y_{\phi}\sum_f x_f\bar{f}f \phi, \label{eq:lagrangian}
\end{equation}
where $\ell=e,\mu$, $F_{\mu\nu}$ ($Z^\prime_{\mu\nu}$) symbolizes the energy-stress tensor of the SM photon ($Z^\prime$) field, $g_i$ denotes the associated coupling, $y_{\phi}$ is the SM fermion flavor-universal coupling, and $x_f$ parameterizes the flavor-nonuniveral Yukawa fractions. For the $\nu$-$\phi$ interaction, one may consider either the Dirac type $\bar{\nu}_\ell\nu_R\phi$ or the Majorana type $\bar\nu_\ell^c\nu_\ell\phi$.
For the light dark matter solution, we use 
an alternative (but similar)
Lagrangian that generates production of dark natter via $Z'$ and/or $\phi$, as well as $\chi \mathcal{N} \to \chi \mathcal{N} \gamma$ for a fermion $\chi$ with mass $m_\chi$:
\begin{equation}
    \resizebox{0.49\textwidth}{!}{$\mathcal{L}_{\chi} \supset g_{\chi} \bar{\chi}\gamma^{\mu}\chi Z^\prime_{\mu} +y_{\chi}\bar{\chi}\chi \phi -\frac{g_{\phi Z^\prime \gamma}}{2}\phi F_{\mu \nu}{Z^\prime}^{\mu \nu} + y_{\phi}\sum_f x_f\bar{f}f \phi$.}
    \label{eq:lagrChi}
\end{equation}

In this scenario, $\chi$ is predominantly produced through exotic three-body decays of charged mesons to explain the anomaly, followed by its decays into $\chi$ pairs. Example UV-complete models realizing the above interactions will be discussed later in the paper.
Based on the above 
Lagrangians, we calculate the differential cross section $d\tilde{\sigma}$ for the $2\to3$ scattering processes, $\nu/\chi + \mathcal{N} \to \nu/\chi + \mathcal{N}+\gamma$, between $\nu/\chi$ and a stationary nucleus or nucleon $\mathcal{N}$ with mass $m_\mathcal{N}$. See Appendix~\ref{app:crosssection2to3} for calculational details.

It is noteworthy that since the typical energy scale of the incoming $\nu/\chi$ that takes part in the $2\rightarrow 3$ process is $\mathcal{O}(0.1-1)$~GeV, and therefore, the energy and angular (cosine) spectra receive contributions from both coherent and incoherent regimes. The former regime is relevant for momentum transfers less than the de Broglie wavelength of the nucleus [$Q^2 \lesssim (100~\text{MeV})^2$], whereas the latter corresponds to momentum transfers of the order of the proton radius. Here $Q^2=-(p_{\mathcal{N},{\rm out}}-p_{\mathcal{N},{\rm in}})^2$. The characteristic momenta range and the associated nucleus/nucleon responses for the coherent and incoherent regimes are imposed using the Helm form factor~\cite{Lewin:1995rx, Dobrich:2015jyk}, $F_{\text{helm}}(Q^2, A)$, and dipole form factor~\cite{Kharzeev:2021qkd}, $F_{\text{dipole}}(Q^2)$, respectively.

In the coherent regime, the Yukawa charge of the nucleus is treated as the coherent sum of the couplings to individual nucleons, whereas in the incoherent regime, nucleons contribute independently. Since the scalar couplings are defined in terms of fundamental quarks in Eq.~\eqref{eq:lagrangian}, they can be related to nucleons using the form factors $f^{p,n}_{Tq}$~\cite{DeRomeri:2024iaw, Cirelli:2013ufw}, which are associated with the spin-independent operators defined in Ref.~\cite{DelNobile:2021wmp}. Therefore, the coherent and incoherent charges for a nucleus with an atomic number and mass $Z$ and $A$, where $N = A-Z$ is the number of neutrons, are,

\begin{equation}
    \begin{aligned}
        C^{S}_{\text{coh}} &= \bigg(Z\sum_{q=u,d} x_q f^p_{Tq}\frac{m_p}{m_q} + N\sum_{q=u,d} x_q f^n_{Tq}\frac{m_n}{m_q}\bigg)^2 \\
        C^{S}_{\text{inc}} &= Z \bigg(\sum_{q=u,d} x_q f^p_{Tq}\frac{m_p}{m_q}\bigg)^2 + N\bigg(\sum_{q=u,d} x_q f^n_{Tq}\frac{m_n}{m_q}\bigg)^2.
    \end{aligned}
    \label{eq:cohincohcharge}
\end{equation}

We account for both the coherent and incoherent cross sections $d\tilde{\sigma}(\text{coherent}) = C^S_{\text{coh}}F^2_{\text{helm}}(Q^2, A)d\tilde{\sigma} (\mathcal{N} = \text{nucleus})$ and $d\tilde{\sigma}(\text{incoherent}) = C^S_{\text{inc}}F^2_{\text{dipole}}(Q^2)d\tilde{\sigma} (\mathcal{N} = \text{nucleon})$. See Appendix~\ref{app:formfactors} for details on the form factors.

\begin{figure*}[t]
    \centering
    \begin{subfigure}{.24\textwidth}
        \includegraphics[width = \textwidth]{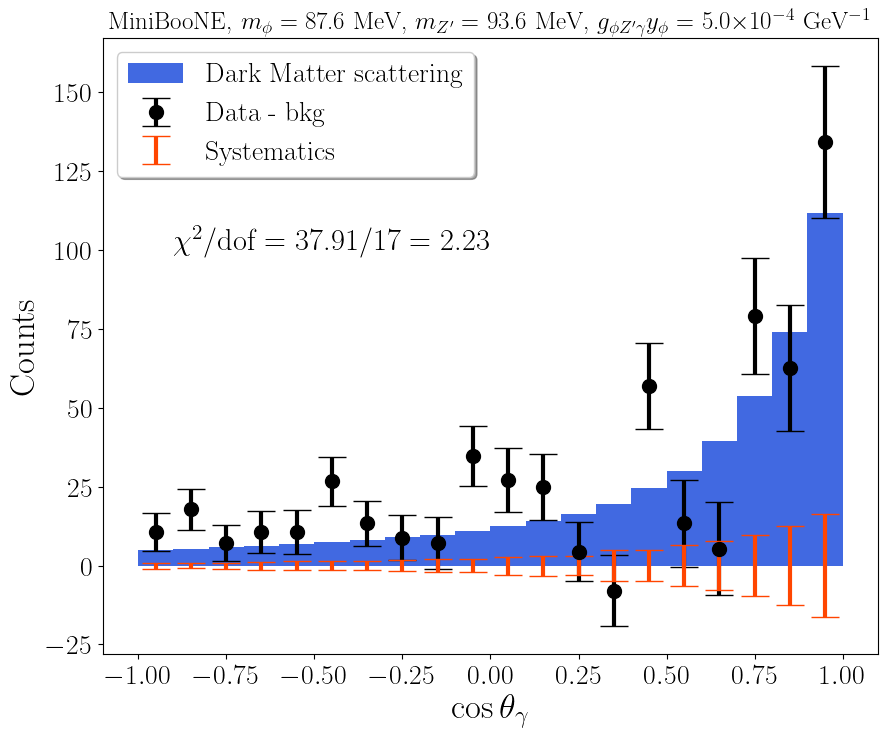}
    \end{subfigure}
    \begin{subfigure}{.24\textwidth}
        \includegraphics[width = \textwidth]{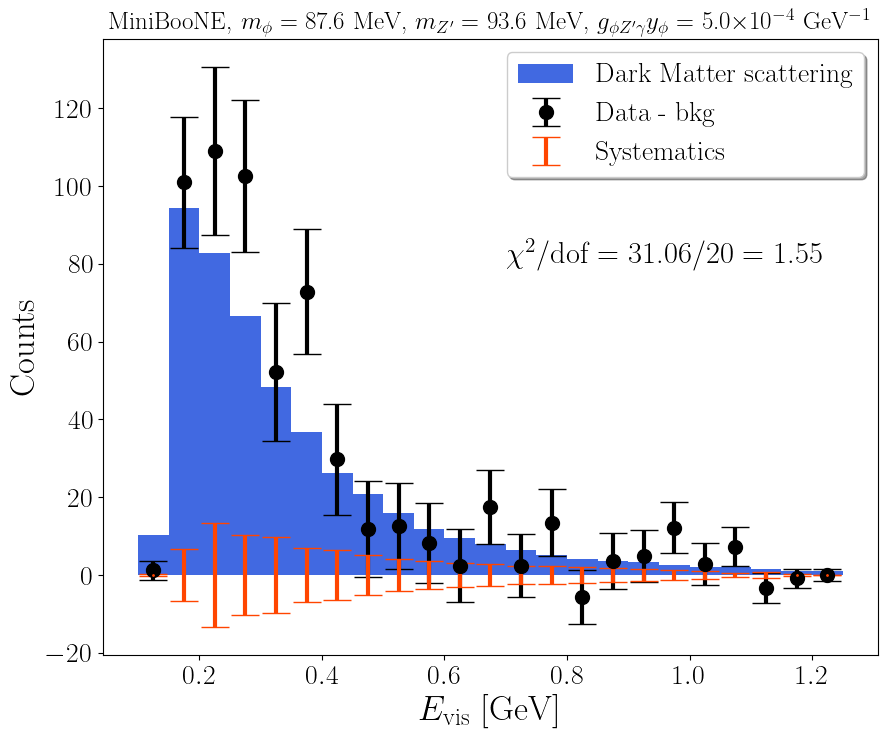}
    \end{subfigure}
    \begin{subfigure}{.24\textwidth}
        \includegraphics[width = \textwidth]{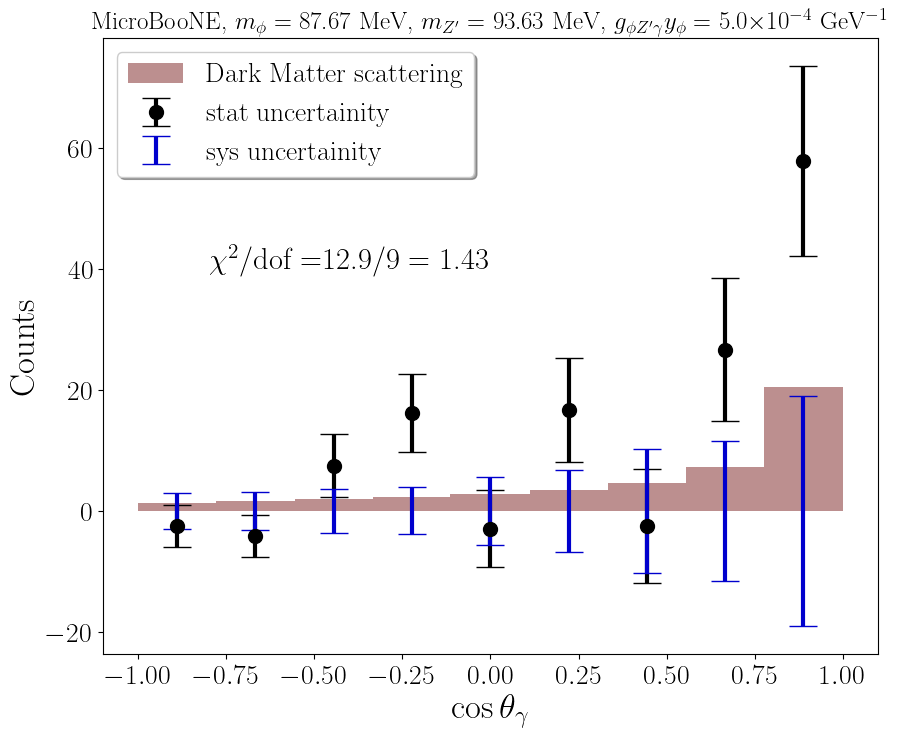}
    \end{subfigure}
    \begin{subfigure}{.24\textwidth}
        \includegraphics[width = \textwidth]{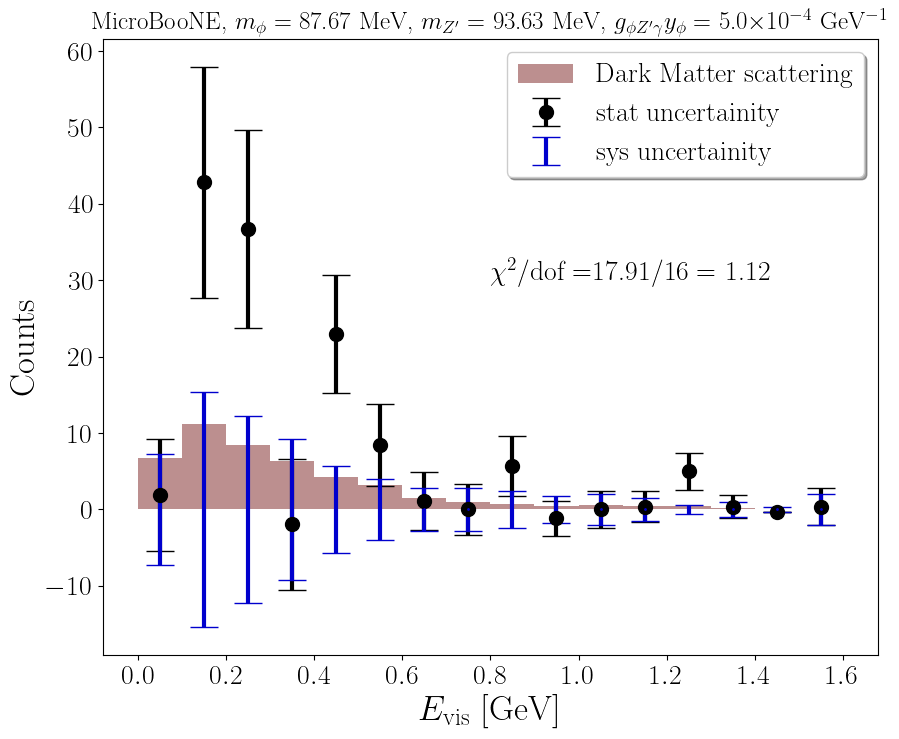}
    \end{subfigure}
    \captionsetup{justification=Justified, singlelinecheck=false}
    \caption{MiniBooNE fits (left) and MicroBooNE overlays (right) with dark matter coupled to a $Z^\prime$ and $\phi$ exchanged with the nucleus where $m_{Z^\prime} = 93.6$~MeV, $m_{\phi} = 87.6$~MeV, $m_\chi = m_{Z^\prime}/3$ and $g_D = 0.5$. After maximizing the dark matter flux from meson decays, the fits require $g = g_{\phi Z'\gamma}y_{\phi} = 5\times 10^{-4}~\text{GeV}^{-1}$. The black and colored error bars depict the statistical and assumed systematic errors, respectively.}
    \label{fig:neufits_sep}
\end{figure*}

\medskip

\noindent {\bf Methodology.}
We begin our analysis by estimating the neutrino/dark matter flux at the BNB. For the neutrino fluxes at the detector, we adopt the published data in Ref.~\cite{MiniBooNE:2008hfu} for MiniBooNE, and Ref.~\cite{MicroBooNE:2016pwy} for MicroBooNE. 
Dark matter particles originate from mediators, which originate from charged mesons, such as $\mathcal{M}^\pm \to \ell^\pm + \nu_l + Z^\prime/\phi$.
Since the magnetic horn system focuses charged mesons before they decay, it is essential to simulate the horn effects accurately. To this end,  ``focused'' charged-meson fluxes are generated using the \texttt{RKHorn} package~\cite{RKHorn2024}. We then simulate their decays into $Z^\prime/\phi$, which subsequently decay into dark matter, collecting the resulting
candidates directed toward the MiniBooNE and MicroBooNE detectors. Since the candidates reaching MiniBooNE and MicroBooNE are highly boosted along the beamline, we assume that they are aligned with the beamline. 
We find that the energy flux of the dark matter reaching the detector is largely independent of the Lorentz structure of the mediator, and the mean energy is generally $\sim$300-400~MeV less than that of the neutrino flux.

We next perform Monte Carlo simulations to estimate the energy and angular spectra of the photon in the $2\to3$ scattering process detailed in the previous section. The technical details for obtaining these kinematic distributions of final-state photons are described in the Appendix~\ref{app:kinematics}. 
Since the cosine spectra of the MiniBooNE and MicroBooNE excesses have a striking forward and off-forward nature, we find that Yukawa couplings must equally accommodate momentum transfers in the coherent and incoherent regimes. This can be ensured if the coherent and incoherent charges in Eq.~\eqref{eq:cohincohcharge} are somewhat at par. 
We accomplish this by
invoking a mild cancelation ($\sim$ 10 -- 30\%) between $x_u$ and $x_d$ in the coherent charge that balances the $Z^2~\rm{or}~N^2$ enhancement.
For illustration, we choose an example of $x_u \sim 1$ and $x_d \sim -1.2$, so that the coherent and incoherent charges have similar magnitudes. 

In our analysis, we compute the single-photon signal from neutrino and dark matter scattering separately. The neutrino fits depend on two mass parameters ($m_{\phi}$ and $m_{Z'}$) and three couplings. However, they can be effectively reduced to three, as the fit is sensitive to the product of couplings, i.e., $g \equiv g_{\nu}g_{\phi Z' \gamma} y_{\phi}$.
In contrast, the dark matter fits additionally depend on $m_\chi$ and $g_\chi$. 
The specific choices of these additional parameters do not affect our main conclusion, although we do keep $m_\chi$ light enough as to not slow down the dark matter propagation outside of the MiniBooNE timing constraint~\cite{MiniBooNE:2020pnu}. For illustration, we adopt the conventional choice $m_\chi=\dfrac{m_{Z^\prime}}{3}$ and $\alpha_\chi \equiv \dfrac{g_\chi^2}{4\pi}=0.5$. Assuming that the mediator production from charged-meson decays is maximized---thereby maximizing the dark matter flux---the dark matter fits depend on three parameters, i.e., $m_{\phi}, m_{Z'},$ and $g\equiv g_{\phi Z' \gamma}y_{\phi}$.
Since more data are available from MiniBooNE's neutrino mode, we scan over the three parameters, $m_{Z^\prime}$, $m_\phi$, and $g$, to find the best-fit values (see Table~\ref{tab:chisq}) by minimizing $(\chi^2/\text{dof})_{\text{cosine}} + (\chi^2/\text{dof})_{\text{energy}}$ with the excess in MiniBooNE's neutrino mode (see Appendix.~\ref{app:chisq}). Using MiniBooNE neutrino mode's best-fit parameters, we find the overlay of the new physics signal with MiniBooNE's antineutrino mode and MicroBooNE neutrino mode excess.

\begin{table}[hb!]
    \centering
    
    \begin{tabular}{|p{0.5cm}|p{3.2cm}|p{1.8cm}|p{1cm}|p{1.3cm}|}
        \hline
        \multicolumn{5}{|c|}{Dark matter fits ($\alpha_{\chi}= 0.5$)} \\
        \hline
        &  & $g_{\phi Z' \gamma}y_{\phi}$ & \multicolumn{2}{|c|}{Average $\chi^2/$dof}\\\cline{4-5}
         No & $(m_{\phi}, m_{Z^\prime}, m_{\chi})$ [MeV] &  [GeV$^{-1}$] & {$\nu$-MB} & {$\mu$B} \\\cline{4-5}
        \hline
        \circled{1} & (45.41, 48.49, 16.17) & $2\times 10^{-4}$ & 2.20 & 1.28\\
        \circled{2} & (87.67, 93.63, 31.21) & $5\times 10^{-4}$ & 1.89 & 1.27\\
        \circled{3} & (119.08, 114.06, 38.02) & $7\times 10^{-4}$ & 2.21 & 1.34 \\
        \circled{4} & (193.07, 138.95, 46.31) & $1\times 10^{-3}$ & 1.79 & 1.41 \\
        \hline
    \end{tabular}
    \vspace{0.3cm} 
    \begin{tabular}{|p{0.5cm}|p{2.4cm}|p{1.8cm}|p{1cm}|p{1cm}|p{1cm}|}
        \hline 
        \multicolumn{6}{|c|}{${\rm Neutrino~fits}$} \\
        \hline
        &   &  & \multicolumn{3}{|c|}{Average $\chi^2/$dof}\\\cline{4-6}
         No & ($m_{\phi},  m_{Z^\prime}$) [MeV] & $g$ [GeV$^{-1}$] & {$\nu$-MB} & {$\bar{\nu}$-MB} & {$\mu$B} \\\cline{4-6}
        \hline
        \circled{1} & (45.41, 48.49) & $5.9\times 10^{-7}$ & 2.02 & 1.49 & 1.13 \\
        \circled{2} & (87.67, 93.63) & $9.0\times 10^{-7}$ & 1.76 & 1.56 & 1.20 \\
        \circled{3} & (119.08, 114.06) & $1.11\times 10^{-6}$ & 1.75 & 1.58 & 1.28 \\
        \circled{4} & (193.07, 138.95) & $1.55\times 10^{-6}$ & 1.75 & 1.66 & 1.35 \\
        \hline
    \end{tabular} 
    \captionsetup{justification=Justified, singlelinecheck=false}
    \caption{Summary of the $\chi^2$ values of the MiniBooNE fits and MicroBooNE overlays for various  $m_\phi$ and $m_{Z^\prime}$ values ($m_{Z^\prime} = 3 m_\chi$). The effective coupling for dark matter fits are given in terms of the couplings required for scattering, i.e., $g_{\phi Z'\gamma}y_\phi$, and the neutrino fits are given by $g = g_\nu g_{\phi Z^\prime \gamma}y_{\phi}$.}
    \label{tab:chisq}
\end{table}

\medskip

\noindent {\bf Results.}
A crucial factor determining the photon signal at MiniBooNE is the relative mass gap between $\phi$ and $Z^\prime$. Although the MiniBooNE excess peaks at a lower energy ($E_{\gamma} \sim 200-300$~MeV), it exhibits a long tail extending up to 1.2 GeV. Similarly, while the MiniBooNE cosine excess peaks at $\cos\theta = 1$, a prominent off-forward component is also observed. Given that neutrinos or dark matter at MiniBooNE peak around 500 MeV, $m_{Z^\prime}$ must be sufficiently large to enable significant energy transfer to the photon, yet not so large that it transfers nearly all of the energy. Additionally, the scalar mediator exchanged between the $Z^\prime$-$\gamma$ system and the nucleus must not be much heavier than the $Z^\prime$; otherwise, the energy would predominantly be transferred to the nucleons or nucleus. At the same time, the scalar must be heavy enough to allow for momentum transfer to the nucleus, facilitating the observed off-forward photon emission.

Figure~\ref{fig:neufits_sep} displays an example MiniBooNE fit along with the MicroBooNE overlay (\circled{2} in Table~\ref{tab:chisq}), 
for 
dark matter scattering. 
We also identify other combinations of $m_{Z^\prime}$, $m_\phi$, and $g$ that provide fits to the energy and cosine excesses observed at MiniBooNE, yielding similar values of $\chi^2$/dof. These parameters are shown in Table~\ref{tab:chisq} along with the average $\chi^2$/dof (averaged over the cosine and energy fits). We find that mediators with comparable masses in the range of $50~\text{MeV} \lesssim m_{\phi/Z'} \lesssim 200~\text{MeV}$ provide good fits to the MiniBooNE excess, with an average $\chi^2$/dof between 1 and 2. Table~\ref{tab:chisq} also contains the fit parameters for neutrino-only scattering. 
While the dark matter flux's mean energy is less than that of the neutrino flux, the energy flux above 600 MeV, which is the main contributor to the $2\to 3$ cross section, has a similar trend. Therefore, within the errors of the measurements, the shape of the neutrino excess is similar to that of dark matter shown in Fig.~\ref{fig:neufits_sep}, consequently leading to qualitatively similar $\chi^2$ values. Furthermore, we find that the same parameter choices yield the same $\chi^2$/dof when fitting the excess by combining neutrino and dark matter signals (see Appendix.~\ref{app:fits}), assuming that dark matter is produced via the same neutrinophilic $Z'$ that facilitates neutrino $2\to 3$ scattering. 
We further observe that the dark matter flux and the $2\to 3$ cross section change only minimally with variations in dark matter mass, as long as $2m_{\chi} < m_{Z^\prime}$. This minimal sensitivity of the $\chi^2$/dof values to $m_{\chi}$ is demonstrated in Appendix.~\ref{app:fits}.

Although the fits were determined by minimizing $\chi^2$ only over MiniBooNE's neutrino-mode data, we find that these same parameters yield $\chi^2$/dof values at MicroBooNE that are lower than $\chi^2$/dof obtained under the null hypothesis, i.e., SM events only hypothesis. For reference, the average $\chi^2/\text{dof}$ for the MiniBooNE neutrino mode, anti-neutrino mode, and MicroBooNE under the SM-only hypothesis are 4.83, 2.02, and 1.69, respectively\footnote{The MiniBooNE values are for our assumed systematics. One must include the covariance matrix to match the official results}. We also find that the new physics signal at MicroBooNE does not reproduce the forward and lower energy excess as effectively as the fits for MiniBooNE. This is mainly due to the choice of flavor-dependent Yukawa couplings, which result in reduced coherent contributions at the MicroBooNE detector. 
However, the $\chi^2$/dof values in MicroBooNE in Table~\ref{tab:chisq} suggest that smaller $m_\phi$ and $m_{Z^\prime}$ values, such as fit \circled{1}, are preferred to reasonably accommodate both MiniBooNE and MicroBooNE excesses under the current Yukawa coupling choices. Given the significantly lower statistics in MicroBooNE compared to MiniBooNE's neutrino mode, we refrain from drawing quantitative conclusions based on the MicroBooNE overlays.

\medskip

\noindent {\bf Constraints.} The recent analysis from MicroBooNE constrains the neutrino flux-averaged coherent cross section to obey $\sigma < 1.49\times 10^{-41}~\text{cm}^2$~\cite{MicroBooNE:2025rsd}. However, the distribution of single photons as a function of true $E_\gamma$ and $\cos\theta_\gamma$
for the $2\to 3$ scattering differs from that used in Ref.~\cite{MicroBooNE:2025rsd}. As shown in Appendix~\ref{app:coherentbounds}, the $2\to 3$ BSM process we consider here gives rise to photons that are more energetic than those from the SM $\text{NC}1\gamma$ process, which arise from $Z$-boson exchange with the nucleus that induces a ground-state-to-ground-state transition in the nucleus with an associated photon~\cite{PhysRevC.89.015503}. Therefore, this cross section constraint only weakly applies, with a $\sim$10\% overlap in the true photon energy and angular distribution template in Ref.~\cite{MicroBooNE:2025rsd} to the coherent part of the $2\to 3$ process. We find that 10\% of the coherent $2\to 3$ cross section satisfies the upper bound of $1.49\times 10^{-41}~\text{cm}^2$.

Existing searches from experiments such as NA62, PIENU, BaBar, CHARM, and E787 would constrain the mass and coupling of the mediators in our model. Since the $2\to 3$ scattering diagram involves an $\mathcal{O}(1)$ coupling between $Z'$ and dark matter, the coupling $g_{Z'}$ is constrained by invisible searches of $Z'$ where $g_{Z'}$ is the coupling of $Z'$ with the SM sector. These searches include electron recoils at LSND, MiniBooNE, nuclear recoils at COHERENT~\cite{COHERENT:2021pvd}, CCM~\cite{CCM:2021leg}, missing energy at NA64~\cite{Banerjee:2019pds}, and monophoton signals at BaBar~\cite{BaBar:2014zli}. Furthermore, the three-body decay constraints on charged pions and kaons at PIENU~\cite{PIENU:2021clt} ($\pi^{\pm} \to l^{\pm} \overline{\nu}_lZ'$) and NA62~\cite{NA62:2021bji} ($K^{\pm} \to \mu^{\pm} \overline{\nu}_\mu Z'$) would also constrain the maximum allowed value of $g_{Z'}$. Similarly, based on the exact Yukawa couplings between the scalar and SM fermions, $y_{\phi}$ is constrained from E787/E949~\cite{E787:2002qfb, E787:2004ovg, E949:2007xyy} on the visible branching ratio of $K^+\to \pi^+ +\phi$, MAMI~\cite{A2atMAMI:2014zdf} on the decay width of $\eta \to \pi^0+(\phi\to \gamma+\gamma)$, as well as PIENU ($\pi^{\pm} \to l^{\pm} \overline{\nu}_l\phi$) and NA62 ($K^{\pm} \to \mu^{\pm} \overline{\nu}_\mu\phi$). 

The relevance of these bounds depends on the specifics of the UV model. For example, if the scalar does not couple directly to dark matter, but only via the $Z'$, constraints from monophoton searches at BaBar and CHARM become relevant, which depend on the mass gap $m_\phi-2m_\chi$ in the decay $\phi\to \gamma
\bar{\chi}\chi$, via an intermediate $Z'$. Certain regions of the parameter space may be subject to supernova constraints (e.g., \cite{Dev:2020eam}), depending on additional model parameters that are not required to account for the anomaly. Additionally, our solutions require higher couplings and $m_{\chi}>10$ MeV; the solutions can bypass limits from supernova. The stellar cooling bounds do not constrain $m_{Z'}$ above 1 MeV (e.g., Refs.~\cite{Hardy:2016kme,Li:2023vpv}). However, astrophysical models have more diverse and larger uncertainties~\cite{Li:2023vpv}. Since our solutions call for mediators and dark matter with masses greater than 10~MeV, constraints from CMB and BBN~\cite{Giovanetti:2021izc} are irrelevant. 

\medskip

\noindent {\bf Example UV complete models.} We now discuss UV-complete $Z'$ models which can lead to the fits in the Table~\ref{tab:chisq}. Since the process in Fig.~\ref{fig:feynman_scattering} requires a new $Z'$ and scalar, we consider a $U(1)_{T3R}$ gauge boson~\cite{Dutta:2019fxn, Dutta:2020scq} that couples to second-generation quarks and third-generation leptons, and a scalar that couples to up, down, strange, and charm quarks. Therefore, the $s$ and $c$ quarks contribute to the dimension-5 coupling $g_{\phi Z' \gamma}$, such that
\begin{equation}
    g_{\phi Z' \gamma} y_{\phi} = \bigg( \frac{g_{T3R} y_{\phi}^2}{400}\bigg)\sum_{c,s}^{P=\{L,R\}}\frac{N_c Q_qx^{\phi}_{q_P}x^{T3R}_{q_P}}{m_q},
\end{equation}
where $g_{Z'}$ is identified as $g_{T3R}$.
The fractional gauge and scalar coupling fractions carried by fermions under $U(1)_{T3R}$ $Z'$ and $\phi$ are denoted by $x_{q_P}^{T3R}$ and $x_{q_P}^\phi$, respectively. In the $U(1)_{T3R}$ case, the fractional gauge charges are $x^{T3R}_{s_R} = 2,~x^{T3R}_{c_R} = -2,~x^{T3R}_{\tau_R} = 2,$ and $x^{T3R}_{N_R} = -2$. This choice results in an anomaly-free $Z'$ model. Here, the $T_{3R}$ quantum numbers do not contribute to the electric charge. The relevant fractional Yukawa couplings are $x^{\phi}_{s} = -1.016,~x^{\phi}_{c} = -0.3,~x^{\phi}_{u} = 1,$ and $x^{\phi}_{d} = -1.2$. This choice strongly suppresses the branching fraction of $\phi \to \gamma \gamma$, satisfying the constraints from MAMI~\cite{A2atMAMI:2014zdf}.

\begin{figure}[t]
    \centering
    \includegraphics[width=0.48\textwidth]{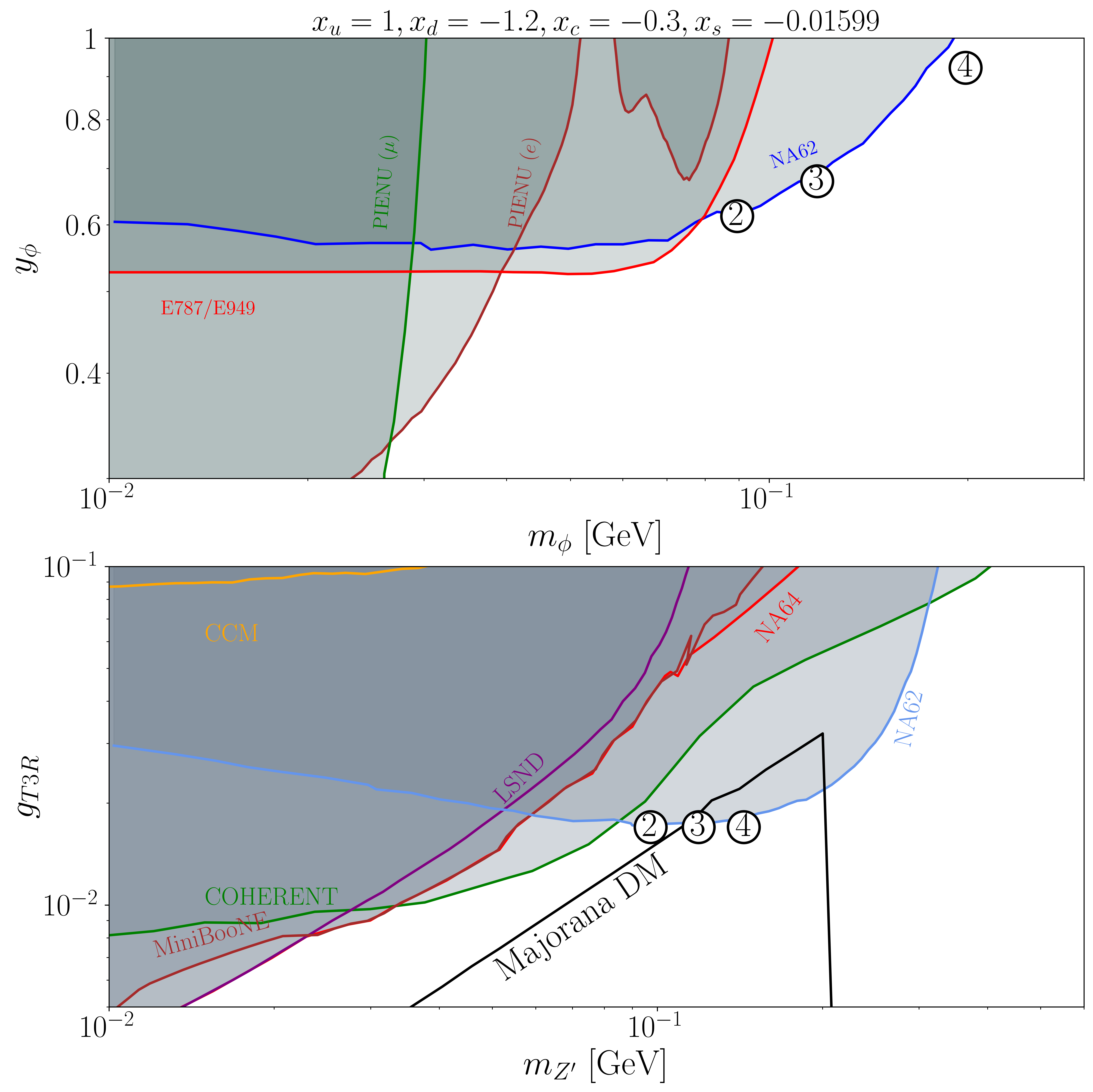}
    \captionsetup{justification=Justified, singlelinecheck=false}
    \caption{ 
    Existing bounds on an invisible $U(1)_{T3R}$ gauge boson $Z'$---i.e., $m_{Z'} > 2m_{\chi}$---that couples to second-generation quarks and third generation-leptons (top), and a scalar coupling to $u,d,s$, and $c$ quarks (bottom).}
    \label{fig:boundsmodel}   
\end{figure}

Figure~\ref {fig:boundsmodel} displays the existing bounds for the gauge boson and scalars for this model, assuming both to decay invisibly. We notice that for $m_{Z'},m_{\phi} >80~\text{MeV}$, the most stringent bounds are set by NA62 that constrains the three-body decay of the charged kaon. Since constraints from $J/\Psi \to \gamma Z'$ decays~\cite{ParticleDataGroup:2024cfk} are more suppressed than NA62 and those from $\Upsilon \to \gamma Z'$ appear only at the loop-level, we refrain from including them in our analysis. Therefore, by adopting the maximum couplings allowed by the NA62 bounds to satisfy the dark matter fits in Table~\ref{tab:chisq}, we simultaneously maximize dark matter production from three-body decays into either $\phi$ or $Z'$. Including the $s$ and $c$ quarks in the loop, we find that the benchmark points \circled{2}, \circled{3}, and \circled{4} satisfy the requirements in Table~\ref{tab:chisq}, for $y_{\phi}$ and $g_{T3R}$ 
marked in Fig.~\ref{fig:boundsmodel}. Note that the couplings simultaneously satisfy the constraints on scattering parameters, i.e., $g_{\phi Z' \gamma}y_{\phi}$, as well as maximize the branching ratio of $K^+ \to \mu^+ +\nu_{\mu}+(Z'/\phi)$, thereby allowing dark matter production to occur via the decays of both the $Z'$ and $\phi$. These scalars can be produced through neutral mesons, via $\pi^{0},\eta\to\gamma \phi$, via operators such as $\partial_\mu \pi^0 \partial_\nu \phi \tilde{F}^{\mu\nu}$ or $\partial_\mu \eta \partial_\nu \phi \tilde{F}^{\mu\nu}$. However, $\partial_\mu \pi^0 \partial_\nu \phi \tilde{F}^{\mu\nu}$ and $\partial_\mu \eta \partial_\nu \phi \tilde{F}^{\mu\nu}$ that are suppressed by a cut-off scale square, i.e., $\frac{1}{\Lambda^2}$. Therefore, constraints from $\pi^0/\eta \to \gamma \phi$ are subdominant.

We note that in the scenario where $\chi$ is predominantly produced from $Z'$, it need not couple to the scalar for our solution, thereby closing the scalar’s invisible decay modes. We discuss the bounds for visibly decaying scalars (i.e., $y_\chi = 0$) in Appendix~\ref{app:vis_scalar_bounds}.

\medskip 
\noindent {\bf Conclusions.}
In this work, we demonstrated that light dark matter and/or neutrino scattering with a nucleus via light mediator exchange can account for the MiniBooNE excess in the $1\gamma0p$ final state without invoking any additional neutrino species. 
These scenarios can be further probed in ongoing experiments such as ICARUS, SBND, and Belle-II. 
Importantly, the $2\to 3$ scattering mechanism plays a crucial role in explaining the excess: vector exchange with the nucleus tends to produce forward-peaked photons, while pseudoscalar exchange leads to off-forward photon emission due to the Lorentz structure of the interaction with the nucleus.

\medskip

\noindent \textbf{Acknowledgements.} We would like to thank Kevin J. Kelly, Rabi Mohapatra, Vishvas Pandey, and Austin Schneider for their helpful discussions. The work of BD and AK is supported by the U.S.~DOE Grant DE-SC0010813. AT acknowledges support in part by the DOE grant DE-SC0010143.  The work of RGV is supported by the Department of Energy Office of Science, High Energy Physics.

\bibliographystyle{bibi}
\bibliography{references}



\appendix

\section{$2\to 3$ cross section}\label{app:crosssection2to3}

We discuss the formulation to calculate the $2\to 3$ cross section for neutrino/dark matter interacting with a nucleon/nucleus. 
The spin-averaged matrix element squared for the $2\to 3$ process $\mathcal{N}(p_a) + \chi/\nu (p_b)  \rightarrow \mathcal{N}(p_1) + \gamma (p_2) + \chi/\nu (p_3)$, can be written in terms of invariants, such that $s_{ij} = (p_i + p_j)^2$ and $t_{ij} = (p_i - p_j)^2$.
\begin{equation}
    \begin{aligned}
        \overline{|\mathcal{M}|^2} =& \frac{(4 m_N^2 - t_{a1})}{(t_{a1} - m_{\phi}^2)^2 (t_{b3} - m_{Z^\prime}^2)^2} \big( 2 m_i^4t_{b3} - 4 m_{i}^2s_{23}t_{b3} \\
        &+ 2 m_i^2 t_{a1}^2 - 2m_{i}^2t_{a1}t_{b3} + 2 s_{23}^2t_{b3} - 2s_{23}t_{a1}t_{b3} \\ & +2s_{23}t_{b3}^2+t_{a1}^2t_{b3}-2t_{a1}t_{b3}^2+t_{b3}^3
        \big)
        \label{eq:matrixelsq}
    \end{aligned}
\end{equation}

To derive the $2\to 3$ scattering cross section, we utilize the following variables and phase-space variables~\cite{Byckling:1971vca}. We first define $\lambda$ and $F$ to represent the kinematic triangular function and flux factor, respectively:
\begin{equation}
    \begin{aligned}
        \lambda (x,y,z) &= (x-y-z)^2 - 4yz, \\ 
        F(s, m_a, m_b) &= 2(2\pi)^{5} \sqrt{\lambda(s,m_a^2, m_b^2)}.
    \end{aligned}
    \label{eq:kin_funcs}
\end{equation}
The phase-space integral denoted by $d\text{PS}_3$ is given by
\begin{equation}
    d\text{PS}_3  = \frac{1}{4\sqrt{\lambda(s,m_a^2, m_b^2)}}\frac{\sqrt{\lambda(s_{2},m_2^2, m_3^2)}}{8s_{23}}d\phi dt_1 ds_2 d\Omega''.
    \label{eq:phaseinteg}
\end{equation}
Here $d\Omega'' \equiv d\cos\theta^{R23}_{b3}d\phi^{R23}_3$ are the angles of particle $3$ defined in the rest frame of $p_2 + p_3$ with the $z$-axis aligned to particle $b$ in the same frame. 
For the remainder of this and the following sections, we denote the angles and other quantities in this frame by a  ``double-prime" superscript, such as $\theta''$ and $\phi''$. We also define several invariant quantities as $s\equiv s_{ab} = (p_a + p_b)^2$, $s_2 \equiv s_{23}= (p_2 + p_3)^2$, and $t_1 \equiv t_{a1} = (p_a - p_1)^2$. 

We then calculate the differential cross section $d\tilde{\sigma}$, using the $2\to 3$ kinematic variables and phase-space elements for a stationary nucleus/nucleon $\mathcal{N}$ target (with mass $m_\mathcal{N}$).
\begin{equation}
    d\tilde{\sigma}\Big|_{\mathcal{N}} = \frac{\overline{|\mathcal{M}|^2} }{F(s, m_\mathcal{N}, m_i)} d\text{PS}_3, 
     \label{eq:diffcs}
 \end{equation}
where $\overline{|\mathcal{M}|^2}$ is the spin-averaged matrix element, calculated in Eq.~\eqref{eq:matrixelsq}. 
The limits of the the five variables $s_2, t_1, \theta'', \phi''$, and $\phi$ are as follows:
\begin{equation}
    \begin{aligned}
        (m_2 + m_3)^2 &\leq s_{23} \leq (\sqrt{s} - m_1)^2,\\
        m_a^2 + m_1^2 -2(E_a^{*}E_1^{*} + p_a^{*}p_1^{*}) &\leq t_{a1} \\
        \leq m_a^2 + m_1^2 &-2 (E_a^{*}E_1^{*} - 2p_a^{*}p_1^{*}), \\
        -1 &\leq \cos{\theta''} \leq 1,\\
        0 &\leq \phi, \phi'' \leq 2\pi,
    \end{aligned}
    \label{eq:kinematicrange}
\end{equation}
where the energy and momenta in the center-of-mass (COM) frame (denoted by the ``star" notation) of the $2\to3$ process are defined as follows:
\begin{equation}
\begin{aligned}
E_a^* &= (s + m_a^2 - m_b^2)/2\sqrt{s}, \\
E_1^* &= (s + m_1^2 - s_{23})/2\sqrt{s}, \\
p_a^* &= \sqrt{E_a^{*2} - m_a^2}, \\
p_1^* &= \sqrt{E_1^{*2} - m_1^2}.
\end{aligned}
\end{equation}

\section{Form factors}\label{app:formfactors}

Given the typical momentum scale through the scalar mediator at MiniBooNE and MircoBooNE, the total cross section of the $2\to 3$ process is given by the mixture between the coherent and incoherent contributions. For momentum transfers smaller than the momentum scale associated with the de Broglie radius of the nucleus, the scattering can occur coherently off the entire nucleus. In this limit, we therefore consider a stationary nuclear scattering target and impose the coherency condition through the Helm form factor~\cite{Lewin:1995rx, Dobrich:2015jyk},
\begin{equation}
    F_{\text{helm}}(t_1, A) = \frac{3j_1(\sqrt{|t_1|}R(A))}{\sqrt{|t_1|}R(A)}e^{-s^2|t_1|/2},
    \label{eq:helmff}
\end{equation}
where $j_1$ denotes the spherical Bessel function, $R(A) = 1.23A^{1/3}~\text{fm}$, and $s = 0.7~\text{fm}$. 

In addition to the coherent contribution, the regime where the scalar interacts with each nucleon is described by the incoherent regime where the momentum transfers are less than the scale of the proton mass. This is incorporated by using the dipole nucleon form factor,
\begin{equation}
    F_{\text{nucleon}} (t_1) = \frac{1}{(1 - t_1/m_{s}^2)},
    \label{eq:dipoleff}
\end{equation}
where $m_s=1.23 \pm 0.07$~GeV~\cite{Kharzeev:2021qkd}.

\section{Kinematics of the single photon from $2\to 3$ scattering}\label{app:kinematics}
For a given incoming neutrino energy or dark matter mass and energy, with fixed $m_{Z^\prime}$ and $m_{\phi}$, we generate weighted scattering events according to the differential cross section given in Eq.~\eqref{eq:diffcs} multiplied with the respective form factors for coherent/incoherent contributions. To accomplish this, we first randomly sample the five variables $s_2, t_1, \theta'', \phi'',$ and $\phi$ within their kinematically allowed ranges as defined in Eq.~\eqref{eq:kinematicrange}. The weight factor is then computed as the product of the phase-space density factor from Eq.~\eqref{eq:phaseinteg} and the matrix element squared.


To determine the total energy and angle of the final-state photon, we first construct the four-momentum of the photon in the rest frame of $(p_2 + p_3)$:
\begin{equation}
    p''_{\gamma} = (E_2'', -p_2''\sin\theta''\cos\phi'', -p_2''\sin\theta''\sin\phi'', -p_2''\cos\theta''),
\end{equation}
where $E''_2 = p_2'' = (s_{23} - m_3^2)/(2\sqrt{s_{23}})$.
It is important to include the negative sign in the three-momentum as $\theta''$ and $\phi''$ are the angles of particle 3 in this frame. Since the photon is particle 2 as per our notation, their direction is opposite to that of particle 3 in the rest frame of $p_2 + p_3$. We now boost the photon to the laboratory frame through two successive Lorentz boost operations. We define a general boost operation by $T(p)$ where $p$ is the four-momentum. The boost factors are obtained from the four-vector via $\gamma = p^0/p$ and $\beta^i = p^i/p^0$ with $p = \sqrt{p^{\mu} p_{\mu}}$.
\begin{enumerate}
    \item $T(-p_{23}^*)$: The first is to boost the photon to the COM frame. This is done by boosting it by the momentum vector $-p_{23}^*$, where $p_{23}=p_2+p_3$ is given by 
    \begin{equation}
        \begin{aligned}
            p_{23}^* = &(E_{23}^*,~p_{23}^*\sin\theta^*_{b3}\cos\phi, \\
            &~p_{23}^*\sin\theta^*_{b3}\sin\phi,~p_{23}^*\cos\theta^*_{a1}),
        \end{aligned}
    \end{equation}
    
    Since $\theta_{b3} = \theta_{a1}$ in the COM frame, the angle is related to the $t_{a1}$ Mandelstam, with $m_a = m_1 = m_\mathcal{N}$, as follows: \begin{equation}
        \cos\theta_{b3} = \cos\theta_{a1} = \frac{t_{a1} - m_a^2-m_1^2 + 2E_a^*E_1^*}{2p_a^*p_1^*}.
    \end{equation}

    \item $T(-p_{ab})$: The second and last step is to boost the photon from the COM frame to the laboratory frame. Since dark matter traveling along the $z$-axis interacts with the stationary nucleus/nucleon $\mathcal{N}$, $p_{ab}=p_a+p_b$ is given by
    \begin{equation}
        p_{ab} = (E_{\chi} + m_\mathcal{N}, 0, 0, p_{\chi}),
    \end{equation}
    where $p_{\chi} = \sqrt{E_{\chi}^2 - m_{\chi}^2}$. 
\end{enumerate}
Therefore, the four-momentum in the laboratory frame is obtained by 
\begin{equation}
    p_{\gamma} = T(-p_{ab})\cdot T(-p_{23}^*)\cdot p''_{\gamma},
\end{equation}
where $p_\gamma$ and $p''_\gamma$ are column vectors and the dots represent the standard matrix multiplications. We then identify $E_{\gamma} = p^0_{\gamma}$ and $\cos\theta_\gamma = p^3_\gamma/p^0_\gamma$.

\section{$\chi^2$ analysis}\label{app:chisq}

Given that there are three degrees of freedom in the fit, the expected number of single-photon events at the detector of interest can be calculated using the following numerical formula:
\begin{widetext}
    \begin{equation}
        \begin{aligned}
            dN^{\chi,{\rm sim}}_{\gamma}( m_{Z^\prime},m_{\phi},g) &= (g_{\phi Z'\gamma}y_{\phi})^2\frac{N_A\rho_T l_T}{A_T}\sum_{E_{i}}N^{\chi}_i \nonumber d \sigma(g_{\chi} = 2.5, g_{\phi Z'\gamma} = 1,y_{\phi} = 1, m_{\chi} = m_{Z^\prime}/3, m_{Z^\prime},m_{\phi}), \\
            dN^{\nu,{\rm sim}}_{\gamma}( m_{Z^\prime},m_{\phi},g) &= (g_{\nu}g_{\phi Z'\gamma}y_{\phi})^2\frac{N_A\rho_T l_T}{A_T}\sum_{E_{i}}N^{\nu}_i d \sigma (g_\nu = 1, g_{\phi Z'\gamma} = 1, y_{\phi} = 1,m_{Z^\prime},m_{\phi}), \\ 
        \end{aligned}
    \label{eq:Simulate}
    \end{equation}
\end{widetext}
where $d\sigma$ implies the total cross section, and where $N_A$ is the number of target nuclei inside the detector fiducial volume, $\rho_T$ and $A_T$ are the density (in g/cm$^3$) and atomic number of the detector material, and $l_T$ is the depth of the detector (in cm). Also, $N_{i}^{\nu}$ and $N_{i}^{\chi}$ are the neutrino and dark matter fluxes reaching the detector, respectively. Here, we assume that the dark matter flux is derived after maximizing the branching ratio of the charged meson decay.

The best-fit parameters are those that minimize the quantity $(\chi^2/\text{dof})_{\text{cosine}} + (\chi^2/\text{dof})_{\text{energy}}$ or the MiniBooNE neutrino-mode excess. Using these best-fit parameters, we then calculate the predicted single-photon events at MicroBooNE. The $\chi^2$ statistic is defined as follows:
\begin{equation}
    \chi^2 = \sum_{i \in \text{bins}} \frac{(d_i - s_i - b_i)^2}{\sigma_{i,\text{stat}}^2+\sigma_{i,\text{sys}}^2},
\end{equation}
where $d_i, s_i$, and $b_i$ are the observed number of events (often referred to as data), the predicted new physics signal, and backgrounds at the $i$th bin, respectively. The parameters $\sigma_{i,\text{stat}}$ and $\sigma_{i,\text{sys}}$ are the statistical and systematic uncertainties respectively. Assuming Poisson statistics, the statistical uncertainty is given by $\sigma_{i,\text{stat}}^2 = d_i$. Since systematic uncertainties for MiniBooNE are available for the reconstructed energy spectrum, we approximate the systematics for both the visible energy and cosine spectra using the total fractional uncertainty reported in Ref.~\cite{MiniBooNE:2020pnu}, which is $f_{\text{sys}} = 3.7\%$. Therefore, the systematic uncertainty is estimated as $\sigma_{i,\text{sys}} = f_{\text{sys}} d_i$. For MicroBooNE, we use the systematic uncertainties provided in Ref.~\cite{MicroBooNE:2025ntu}. 

\section{Fits} \label{app:fits}

\begin{table}[h]
    \centering
    \begin{tabular}{|p{0.5cm}|p{3.2cm}|p{1.8cm}|p{1cm}|p{1cm}|}
        \hline
        \multicolumn{5}{|c|}{Dark matter fits ($\alpha_{\chi} = 0.5$)} \\
        \hline
        &  &  $g_{\phi Z^\prime \gamma}y_{\phi}$ & \multicolumn{2}{|c|}{$\chi^2/$dof}\\\cline{4-5}
         No & $(m_{\phi}, m_{Z^\prime}, m_{\chi})$ [MeV] &  [GeV$^{-1}$] & {$\nu$-MB} & {$\mu$B} \\\cline{4-5}
        \hline
        1. & (45.41, 48.49, 16.17) & $2\times 10^{-4}$ & 2.20 & 1.28\\
        2. & (87.67, 93.63, 31.21) & $5\times 10^{-4}$ & 1.89 & 1.27\\
        3. & (119.08, 114.06, 38.02) & $7\times 10^{-4}$ & 2.21 & 1.34 \\
        4. & (193.07, 138.95, 46.31) & $1\times 10^{-3}$ & 1.79 & 1.41 \\
        \hline
        5. & (45.41, 48.49, 20.64) & $2\times 10^{-4}$ & 2.23 & 1.26\\
        6. & (87.67, 93.63, 39.8) & $5\times 10^{-4}$ & 1.94 & 1.28\\
        7. & (119.08, 114.06, 51.85) & $ 7\times 10^{-4}$ & 1.88 & 1.35 \\
        8. & (193.07, 138.95, 87.7) & $1\times 10^{-3}$ & 1.80 & 1.40 \\
        \hline
    \end{tabular}
    \captionsetup{justification=Justified, singlelinecheck=false}
    \caption{Summary of $\chi^2$ values for the MiniBooNE fits and MicroBooNE overlays for various values of $m_\phi$ and $m_{Z^\prime}$ for dark matter scattering only (with mass $m_\chi$). The two table sections correspond to different ratios of $m_{\chi}/m_{Z'}$. }
    \label{tab:chisqdm}
\end{table}

\begin{figure*}[ht!]
    \centering
    \includegraphics[width=0.8\linewidth]{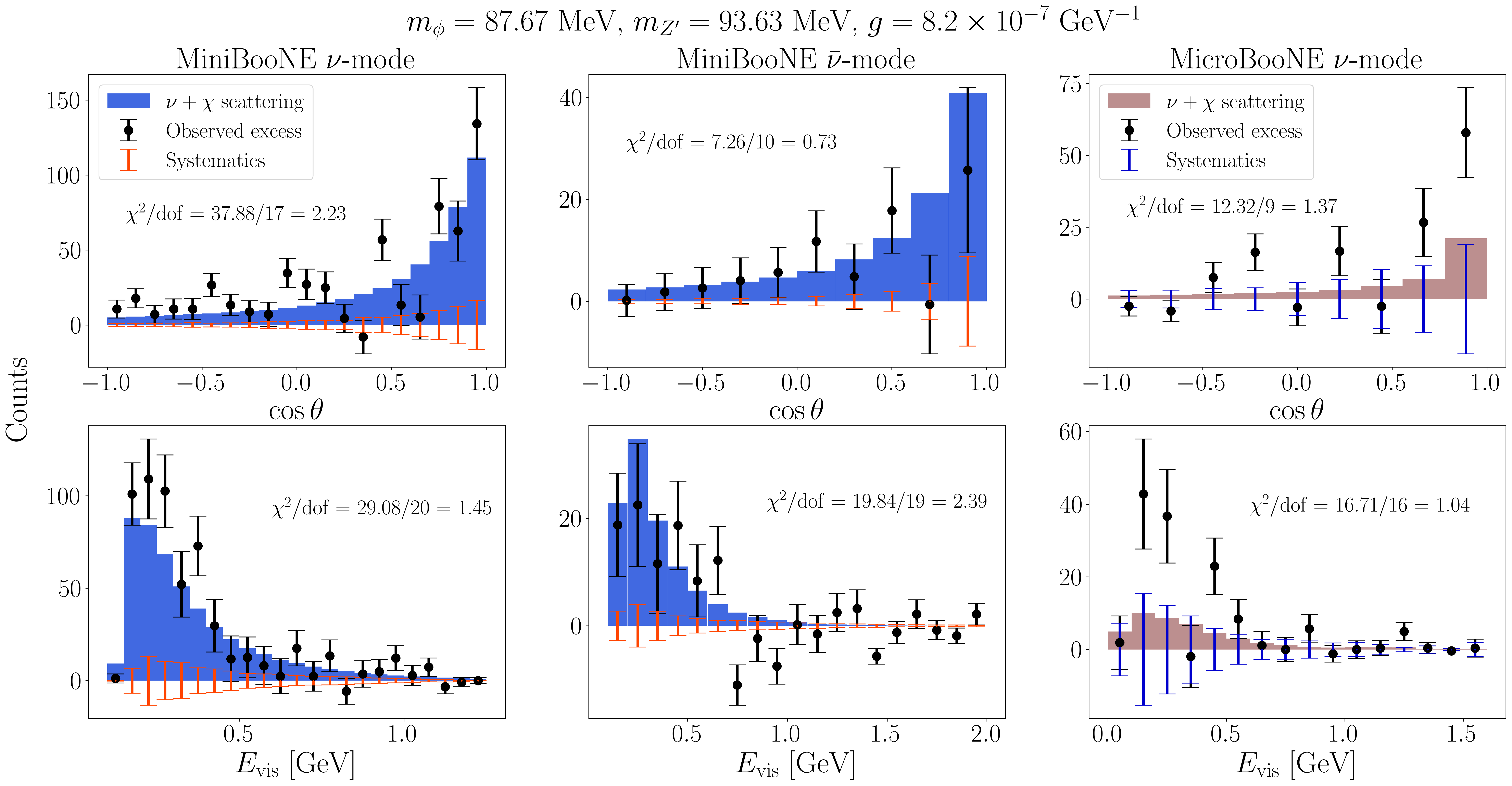}
    \captionsetup{justification=Justified, singlelinecheck=false}
    \caption{Fits to the MiniBooNE neutrino-mode (left), and overlays with anti-neutrino mode (middle), and the inclusive single photon MicroBooNE (right) cosine (upper) and visible energy (lower) excesses for ${\rm neutrino}+{\rm dark}$ matter scattering where the mediators used are $m_{Z^\prime} = 93.6$~MeV, and $m_{\phi} = 87.6$~MeV, and the combination of couplings are $g = g_{\nu} g_{\phi Z'\gamma} y_{\phi} = 8.2\times 10^{-7}~\text{GeV}^{-1}$. The black data points+error bars denote the statistical errors, and the colored error bars correspond to the systematic uncertainties.}
    \label{fig:neudmfits}
\end{figure*}

Figure~\ref{fig:neudmfits} shows the MiniBooNE neutrino mode fits and overlays for the MiniBooNE antineutrino mode and MicroBooNE neutrino mode with the combined neutrino+dark matter signals. We observe that the fits closely resemble the individual fits presented in the paper. Additionally, we report the $\chi^2$/dof values for the dark matter only fits, along with the required effective coupling values, assuming $\alpha_\chi = 0.5$, in Table~\ref{tab:chisqdm}. We further find that the fits are not particularly sensitive to the choice of dark matter mass. This indicates that the dominant factors influencing the fits and the values of $\chi^2$ / dof are the masses of the scalar and vector bosons.

\section{Coherent contribution}\label{app:coherentbounds}

\begin{figure}[h]
    \includegraphics[width=0.45\textwidth]{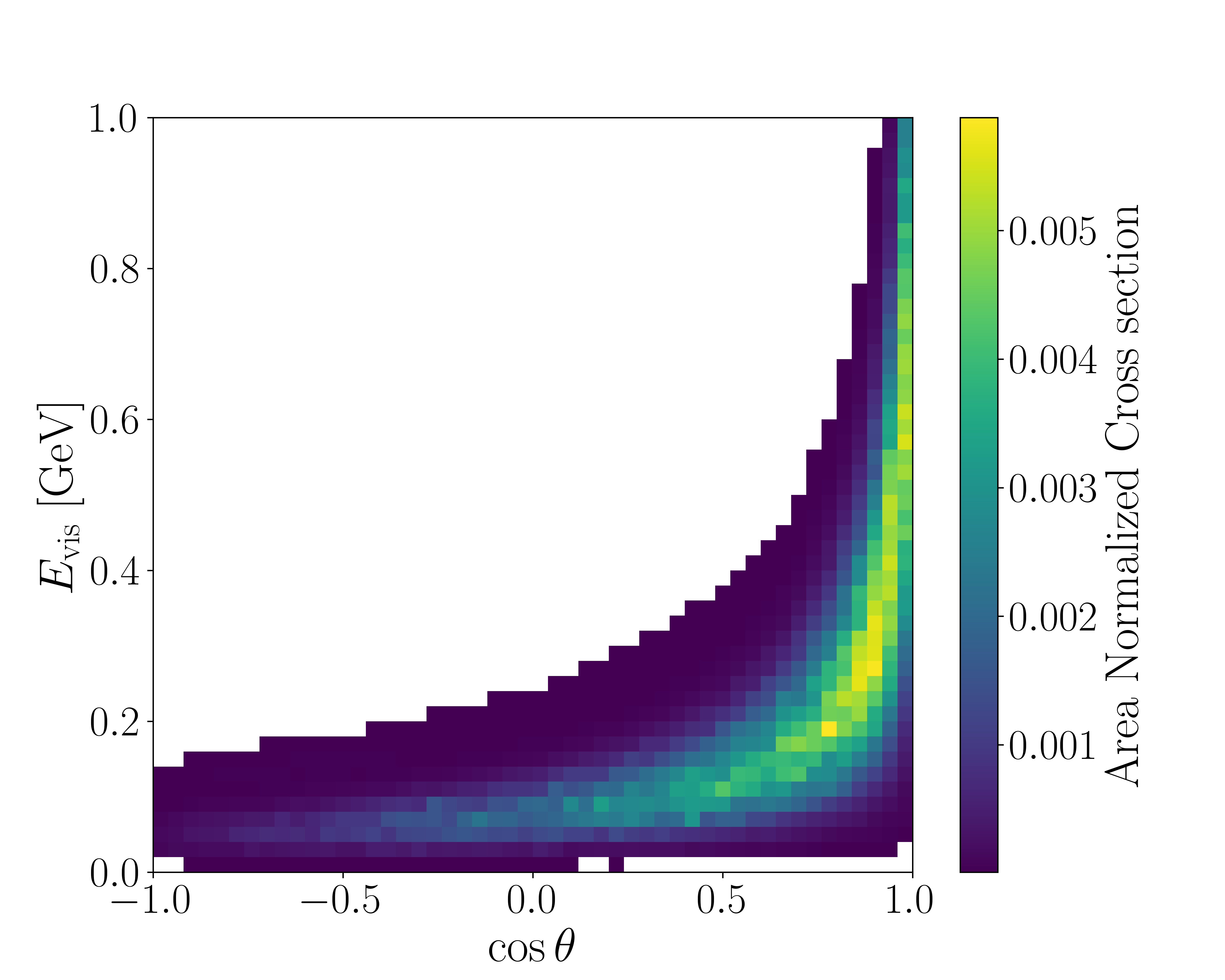}
    \captionsetup{justification=Justified, singlelinecheck=false}
    \caption{The energy cosine distribution of the coherent $2\to3$ process from neutrinos and dark matter for $m_{\phi} = 87~\text{MeV},~m_{Z'} = 93~\text{MeV}$, and $g = 8.2\times 10^{-7}~\text{GeV}^{-1}$.}
    \label{fig:coherent}
\end{figure}

Figure~\ref{fig:coherent} depicts the energy-cosine distribution of the photons from the coherent part of the $2\to3$ process, where the average cross section is $\sigma_{\text{coh}}^{\text{avg}} \simeq 1\times 10^{-40}~\text{cm}^2$. This includes contributions from both neutrinos and dark matter. We see that this differs from the template used in Ref.~\cite{MicroBooNE:2025rsd}, with about $\sim 10\%$ overlap between them. Including this fact, we see that the cross section satisfies the upper bound on the coherent cross section.

\begin{figure}[h]
    \centering
    \includegraphics[width=0.48\textwidth]{bounds_scalar_vis.png}
    \captionsetup{justification=Justified, singlelinecheck=false}
    \caption{Bounds on $y_{\phi}$ as a function of $m_\phi$ for a visible scalar, assuming $g_{Z'} = 2\times 10^{-2}$, $x_u = 1, x_d = -1.2, x_c = -0.3,$ and $x_s = -1.016$. from BaBar and CHARM monophoton searches, for different mass splittings $\Delta m$ (see text for more details). The green dashed line shows the predictions for Belle-II at 50~ab$^{-1}$.}
    \label{fig:boundsgzpg}
\end{figure}

\section{Bounds on visible decaying scalars}\label{app:vis_scalar_bounds}

Figure~\ref{fig:boundsgzpg} depicts the bounds on the Yukawa coupling $y_\phi$ for a visibly decaying scalar. If $y_{\chi} = 0$, i.e., the scalar does not couple to dark matter, the scalar is rendered visible and can produce a single photon via $\phi \to Z' \gamma$/ $\phi \to \bar{\chi}\chi \gamma$. For a visibly decaying scalar, bounds from CHARM are also applied. The decay width of a scalar is larger for a larger mass gap, where we define $\Delta m = m_{\phi}-m_{Z'}$ or $m_{\phi} - 2m_{\chi}$. We find that the bounds shift towards lower Yukawas for larger mass gaps due to lower lifetimes for larger mass gaps. Additionally, we see that the bounds from E949/E787~\cite{E787:2002qfb, E787:2004ovg, E949:2007xyy, BNL-E949:2009dza} have an additional upper limit, as larger couplings result in decays inside the detector, which does not render the scalar as ``invisible".

\end{document}